\documentstyle[eqsecnum,aps,epsf]{revtex}
 
\begin{document}
\title{Spherically symmetric static solution for colliding null dust}
\author{L\'{a}szl\'{o} \'{A}. Gergely}
\address{KFKI Research Institute for Particle and Nuclear Physics,
Budapest 114,\\
P.O.Box 49, H-1525 Hungary}
\maketitle
 
\begin{abstract}
The Einstein equations are completely integrated in the presence of two
(incoming and outgoing) streams of null dust, under the assumptions
of spherical
symmetry and staticity. The solution is also written in double
null and in radiation coordinates and it is
reinterpreted as an anisotropic fluid. Interior matching with a
static fluid and exterior matching with the Vaidya solution along
null hypersurfaces is discussed. The connection with
two-dimensional dilaton gravity is established.
\end{abstract}
 
\section{Introduction}
 
Null dust represents the high frequency (geometrical
optics) approximation to the unidirectional radial flow of
unpolarized radiation. This is a reasonable approximation whenever
the wavelength of the radiation is negligible compared to the
curvature radius of the background.
Various exact solutions of the Einstein field equations
were found in the presence of pure null dust (for reviews see
\cite{Kramer1}, and more recently \cite{Kramer2} and \cite{KuBi}).
 
In some scenarios even gravitation behaves as null dust. Price
\cite{Price1,Price2} has shown that a collapsing spheroid
radiates away all of its initial characteristics excepting its
mass, angular momentum and charge. (This result is known as the
{\it no hair} theorem.) The escaping radiation then interacts
with the curvature of the background being partially backscattered.
Both the escaping and the backscattered radiation can be modeled
by null dust \cite{PI} as the curvature radius of the background is larger
than the wavelength of the radiation.
 
Letelier has shown that the matter source composed of two null
dust clouds can be interpreted as an anisotropic fluid
\cite{Let1},
giving also the general solution for plane-symmetric anisotropic
fluid with two null dust components. Later Letelier and Wang
\cite{Let2} have discussed the collision of cylindrical null dust
clouds.
The collision of spherical null dust streams was discussed by
Poisson and
Israel \cite{PI}. Their analysis yielded the phenomenon of mass
inflation.
 
However, no exact solution in the presence of two colliding null
dust streams with spherical symmetry was known until now. Recently
Date
\cite{Date} tackled this problem under the assumption of staticity,
integrating part of the Einstein equations.
It is the purpose of the present paper to present the
exact solution for the case of two colliding spherically
symmetric null dust streams in equilibrium, in a completely
integrated form.
 
The plan of the paper is as follows. In Sec. II we derive the
field equations and we integrate them. The emerging exact solution
is written explicitly in suitable coordinates adapted to spherical
symmetry and staticity. The metric in radiation
coordinates and in double null coordinates
is also given. We analyze the metric both analytically and by
numeric plots. In Sec. III we present various possible
interpretations of the solution, including the anisotropic fluid
picture, and dilatonic gravity.
 
Finally Sec. IV contains the analysis of the interior and
exterior matching conditions on junctions along timelike and null
hypersurfaces, respectively. We employ the matching procedure of
Barrab\`{e}s and Israel \cite{BI}. The
interior junction fixes the parameters of the solution in terms
of physical characteristics of the central star:
the mass and energy density on the junction.
The exterior matching with incoming and outgoing Vaidya solutions
\cite{Vaidya}
leads to the conclusion that no distributional matter is present
at the junction. This feature is in contrast with the exterior matching
with the Schwarzschild solution \cite{Date}.
 
\section{Solution of the Einstein Equations}
 
The general form of a spherically symmetric, static metric
in a spacetime with topology $R^{2}\times S^{2}$ is
 
\begin{equation}
ds^{2}=-h(r)dt^{2}+f(r)^{-1}dr^{2}+r^{2}d\Omega ^{2}.  \label{metric}
\end{equation}
Here $t$ is time, $r$ is the curvature coordinate (e.g. the radius of the
sphere $t$=const with area $4\pi r^{2}$), $d\Omega ^{2}=d\theta
^{2}+\sin
^{2}(\theta )d\phi ^{2}$ is the square of the solid angle element. The
functions $f(r)$ and $h(r)$ are positive valued. We introduce
the local mass function $m(r)$, related to the gravitational
energy within the sphere of radius $r$ \cite {Sy}:
 
\begin{equation}
f(r)=1-\frac{2m(r)}{r}\ .  \label{mass}
\end{equation}
 
The energy-momentum tensor in the region of the cross-flowing null dust is
a superposition of the energy-momentum tensors of the incoming and outgoing
components:
 
\begin{equation}
T^{ab}=\frac{\beta (r)}{8\pi r^{2}}(v^{a}v^{b}+u^{a}u^{b}).  \label{en-mom}
\end{equation}
All energy conditions are satisfied for $\beta (r)\geq 0$. The
same linear mass density function $\beta $ was chosen for both
components as staticity requires no net flow in either of the null
directions. The vector fields
 
\begin{equation}
v^{a}=\frac{1}{\sqrt{2}}\left( \frac{1}{\sqrt{h}},\sqrt{f},0,0\right)
,\qquad u^{a}=\frac{1}{\sqrt{2}}\left( \frac{1}{\sqrt{h}},-\sqrt{f}
,0,0\right)   \label{vurt}
\end{equation}
are the tangents to the (future-oriented) outgoing and incoming null
congruences, partially normalized such that $v^{a}u_{a}=-1$.
Similarly as for the one-component null dust, here
both null congruences are geodesics \cite{Date}.
 
After eliminating second time derivatives from the nontrivial Einstein
equations we have the system:
\begin{eqnarray}
r\frac{df}{dr} &=&-\beta -f+1  \nonumber \\
rf\frac{dh}{dr} &=&h(\beta -f+1)  \label{sys1} \\
rf\frac{d\beta }{dr} &=&\beta (-\beta +f-1) \ . \nonumber
\end{eqnarray}
 
The solutions with $\beta $=const all reduce to $\beta =0$,
meaning vacuum.
They are either the Schwarzschild or the flat solution, in accordance with
the Birkhoff theorem. For $\beta \neq $const one of the equations is a
relation between the mass density $\beta (r)$ and the metric function $
h\left( r\right) $:
 
\begin{equation}
h=\frac{A}{\beta },  \label{g}
\end{equation}
where $A>0$ is a constant. This relation can be deduced also from the
energy-momentum conservation\cite{Date}. The remaining equations do not
contain the constant $A$:
\begin{eqnarray}
f\frac{d\ln f}{d\ln r} &=&-\beta -f+1  \label{eq1} \\
f\frac{d\ln \beta }{d\ln r} &=&-\beta +f-1 \ . \label{eq2}
\end{eqnarray}
Inserting (\ref{mass}) in (\ref{eq1}) the mass is found to
increase with the radius: $dm/d r=\beta /2>0 $.
 
Now we solve the system (\ref{eq1}),(\ref{eq2}). Following Date\cite{Date} ,
we eliminate $f$ from (\ref{eq1}) by its expression taken from (\ref{eq2}):
 
\begin{equation}
f=\frac{1+\beta }{1-\frac{d\ln \beta }{d\ln r}}.  \label{f}
\end{equation}
The resulting second order ordinary differential equation in $1/\beta $ can
be integrated, finding:
 
\begin{equation}
\frac{d}{d\ln r}\left( \frac{1}{\beta }\right) =\left( D-\frac{1}{\beta }
-2\ln \frac{\beta }{r}\right) \frac{1}{1+\beta }.
\end{equation}
Here $D$ is an integration constant. Inserting this expression in (\ref{f}),
an algebraic relation between the metric components emerges:
 
\begin{equation}
\frac{(1+\beta )^{2}}{2f\beta }=\ln \left( \frac{Cr}{\beta }\right) ,
\label{int1}
\end{equation}
where $C=\exp ((1+D)/2)>0$.
 
Then we complete the integration of the system
(\ref{eq1}),(\ref{eq2}). The key remark is that none of the equations
in this system contains explicitly the independent variable $\ln r$.
Thus one can pass to the new independent variable $\beta $, in terms
of which an ordinary first order equation can be written:
 
\begin{equation}
\frac{df}{d\beta }=\frac{f(\beta +f-1)}{\beta (\beta -f+1)} \ .
\label{eq3} \end{equation}
 
Introducing the new positive variables
 
\begin{equation}
P=(2f\beta )^{1/2},\ \qquad L=\frac{1+\beta }{(2f\beta )^{1/2}}
\label{newvar}
\end{equation}
the equation (\ref{eq3}) takes the form of a first order linear
(inhomogeneous) ordinary differential equation:
 
\begin{equation}
\frac{dP}{dL}=2\left( 1-PL\right) \ . \label{eq4}
\end{equation}
 
The solution is found by integrating first the homogeneous equation, then
varying the constant. It is
 
\begin{equation}
P=2e^{-L^{2}}\Phi _{B}(L)\;,\qquad
\Phi_{B}(L)=B+\int^{L}e^{x^{2}}dx>0, \label{int2}
\end{equation}
where $B$ is a third integration constant. The function $\Phi _{B}(L)$
in (\ref{int2}) can be expressed either in terms of the error function or in
terms of the Dawson function:
 
\begin{equation}
\Phi _{B}(L)-B=-\frac{i\sqrt{\pi }}{2}
\mathop{\rm erf}
(iL)=e^{L^{2}}Dawson(L).
\end{equation}
For properties of these transcendental functions see\cite{Erdelyi}.
 
From Eqs. (\ref{int1}),(\ref{newvar}) and (\ref{int2}) both the
curvature
coordinate $r$ and the metric functions $\beta $ and $f$ are
found as functions of the radial variable $L$:
\begin{eqnarray}
Cr(L) &=&-e^{L^{2}}+2L\Phi _{B}(L)  \label{rL} \\
\beta (L) &=&-1+2Le^{-L^{2}}\Phi _{B}(L)  \label{beL} \\
f(L) &=&\frac{2\Phi _{B}^{2}(L)}{e^{L^{2}}\left[ -e^{L^{2}}+2L\Phi
_{B}(L)\right] }.  \label{fL}
\end{eqnarray}
 
Then the mass function $m=m\left( L\right) $ is obtained from (\ref{mass}):
\begin{equation}
2Cm(L)=-e^{L^{2}}+2L\Phi _{B}(L)-2e^{-L^{2}}\Phi _{B}^{2}(L).
\label{mL} \end{equation}
 
It is easy to check that both $r$ and $m$ are monotonously increasing
functions of $L$:
\begin{eqnarray}
\frac{d\left( Cr\right) }{dL}= &&2\Phi _{B}(L)>0  \label{drL} \\
\frac{d\left( Cm\right) }{dL}=e^{-L^{2}}\Phi _{B}(L) &&\left[
-e^{L^{2}}+2L\Phi _{B}(L)\right] \geq 0.  \label{dmL}
\end{eqnarray}
In the last relation the equality holds for $r=0$.
 
Now we have everything together to write the metric in terms of the new
radial coordinate $L$:
\begin{equation}
ds^{2}=\frac{-Ae^{L^{2}}dt^{2}}{-e^{L^{2}}+2L\Phi
_{B}(L)}+2e^{L^{2}}\left[
-e^{L^{2}}+2L\Phi _{B}(L)\right] \frac{dL^{2}}{C^{2}}+\left[
-e^{L^{2}}+2L\Phi _{B}(L)\right] ^{2}\frac{d\Omega ^{2}}{C^{2}}.
\label{metrictL}
\end{equation}
 
There are three parameters in the solution, two of them restricted to be
positive: $A$ and $C$. Without loss of generality we can choose $A=1$
by rescaling the time coordinate. The parameter $C$ provides some
distance scale. We comment on the third parameter in what follows. Both $
Cr\geq 0$ and the energy conditions $\beta \geq 0$ imply
\begin{equation}
B\geq \chi (L)\ ,\quad \chi (L)=\frac{e^{L^{2}}}{2L}-\int^{L}e^{x^{2}}dx
\label{BL}
\end{equation}
valid for all admissible values of $L$. The equality $B=\chi (L_{0})$ holds
for $L=L_{0}$ corresponding to $r=0$. As $d\chi/dL=-e^{L^{2}}/2L^2<0$, the
function $\chi (L)$ is monotonously decreasing and the inequality (\ref{BL})
will be satisfied for any $L>L_{0}$. Thus $B$ gives the lower boundary $L_{0
\text{ }}$for the range of the radial coordinate $L$.
 
Next we plot numerically the functions $\beta (r)$, $f(r)$ and $
m\left( r\right) $ for different values of the parameter $B$ (Figs.1-3).
\begin{figure}[tbh]
\hspace*{.2in}
\special{hscale=30 vscale=30 hoffset=-20.0 voffset=20.0
         angle=-90.0 psfile=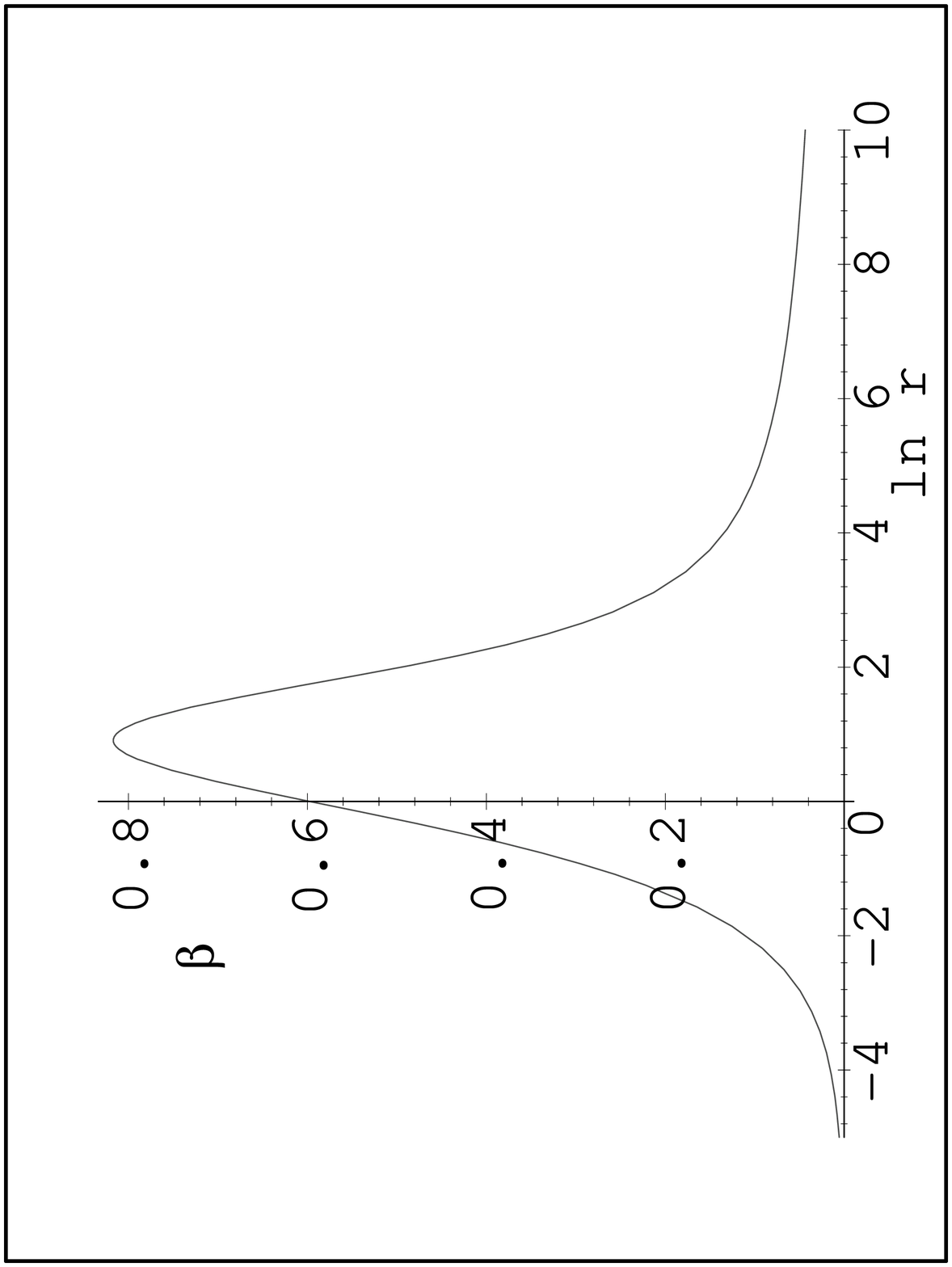}
\hspace*{3.2in}
\special{hscale=30 vscale=30 hoffset=-20.0 voffset=20.0
         angle=-90.0 psfile=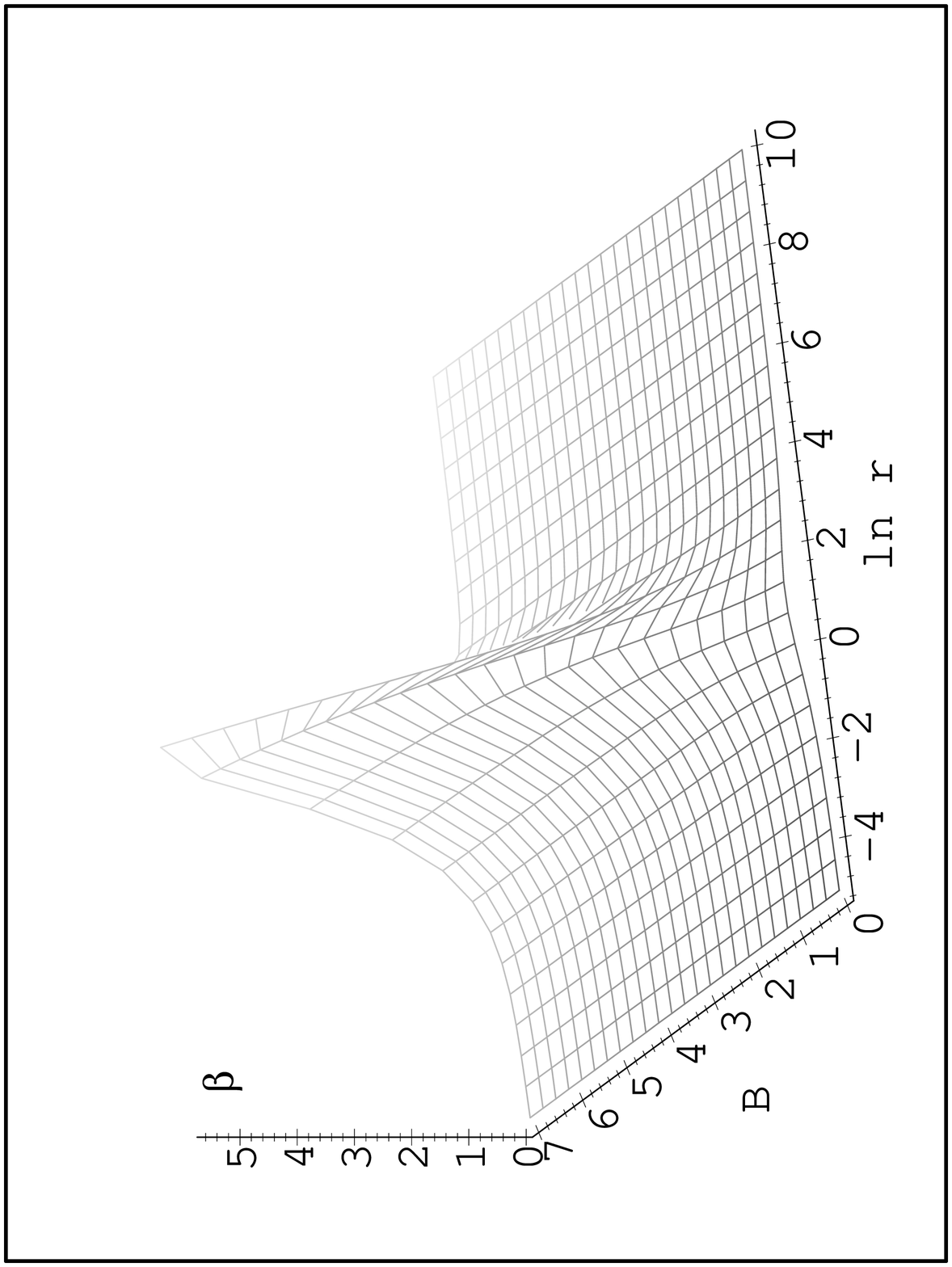}
\vspace*{2.3in}
\caption{(a) The function $\beta =\beta (\ln r)$ at parameter values
$B=C=1$. The metric is singular at $r=0$ and the space-time is
not asymptotically flat. (b)
Plot of $\beta =\beta (\ln r,B)$ for $C=1$ and $0\leq B\leq 7$.}
\end{figure}
\begin{figure}[tbh]
\hspace*{.2in}
\special{hscale=30 vscale=30 hoffset=-20.0 voffset=20.0
         angle=-90.0 psfile=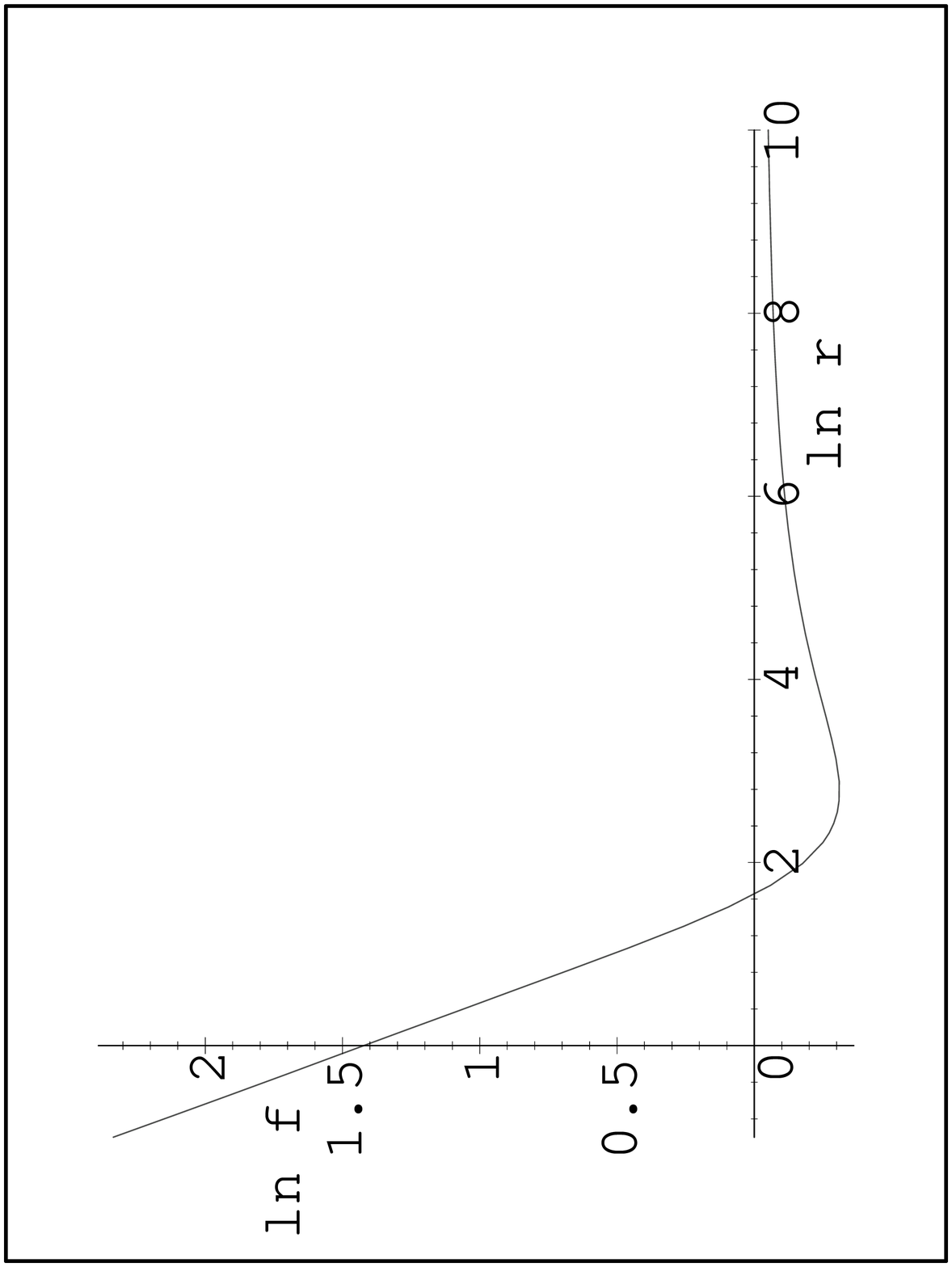}
\hspace*{3.2in}
\special{hscale=30 vscale=30 hoffset=-20.0 voffset=20.0
         angle=-90.0 psfile=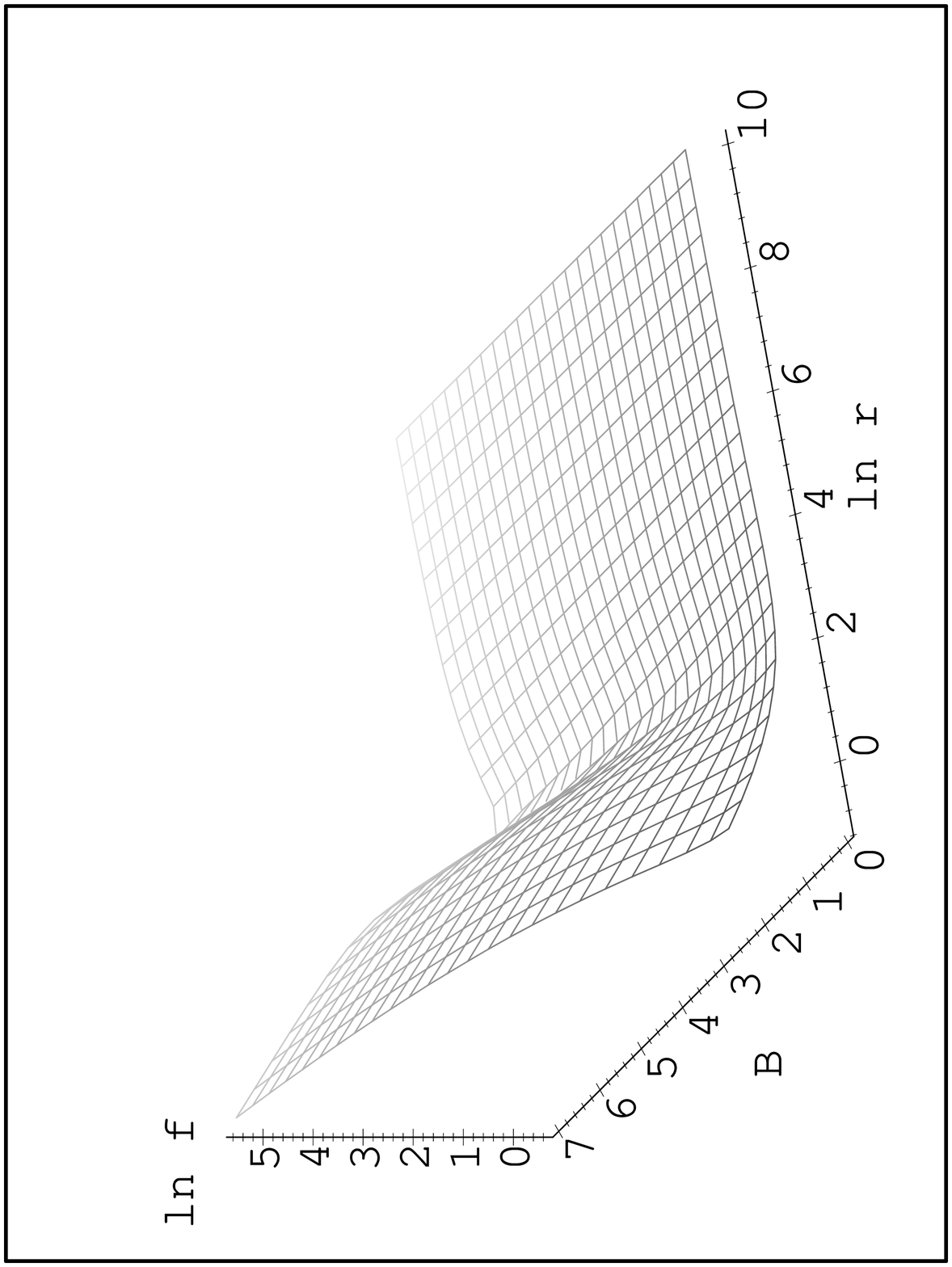}
\vspace*{2.3in}
\caption{(a) The function $\ln f=\ln f(\ln r)$ at parameter values $B=C=1$.
(b) Same for $0\leq B\leq 7$.}
\end{figure}
\begin{figure}[tbh]
\hspace*{.2in}
\special{hscale=30 vscale=30 hoffset=-20.0 voffset=20.0
         angle=-90.0 psfile=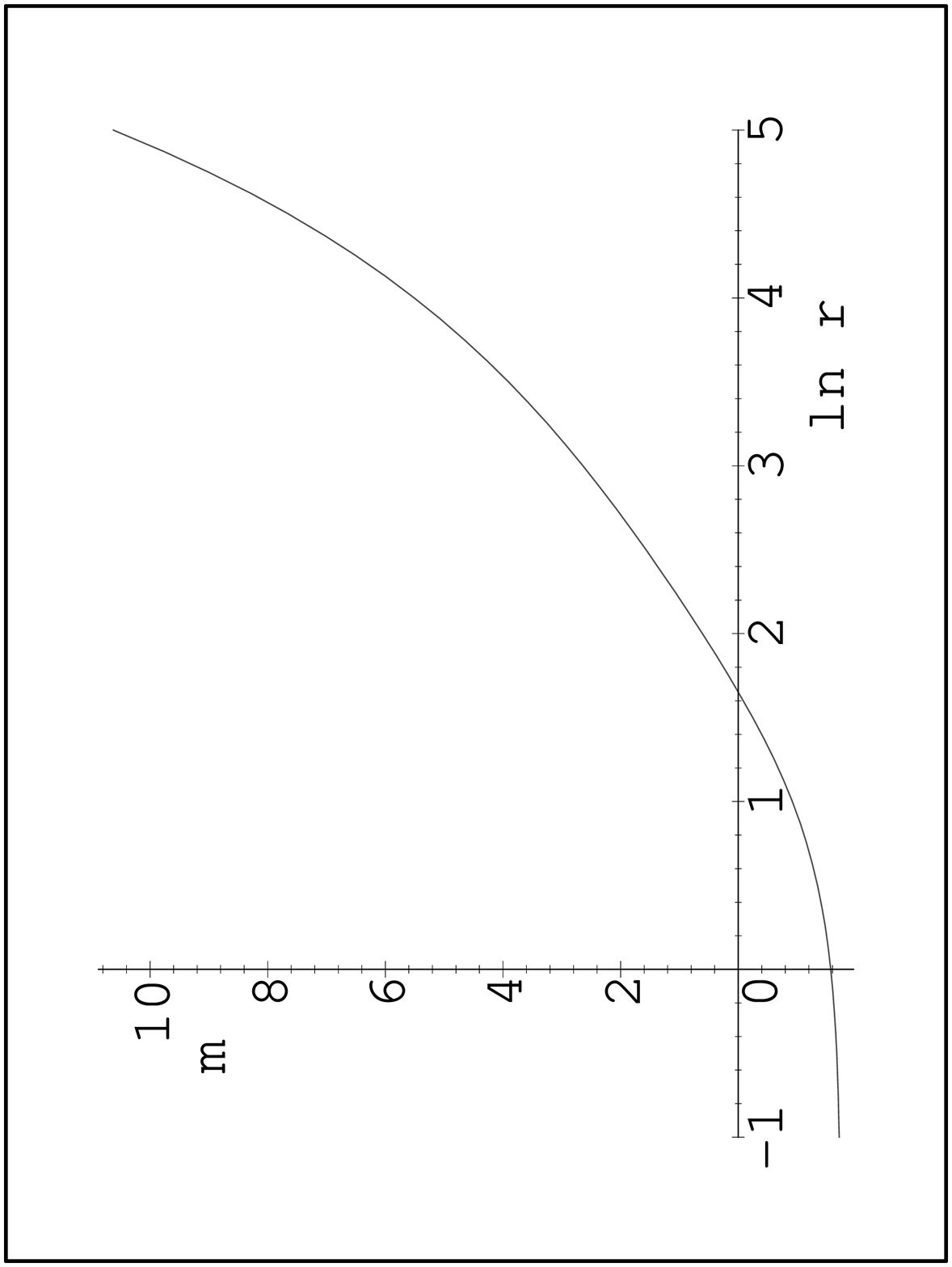}
\hspace*{3.2in}
\special{hscale=30 vscale=30 hoffset=-20.0 voffset=20.0
         angle=-90.0 psfile=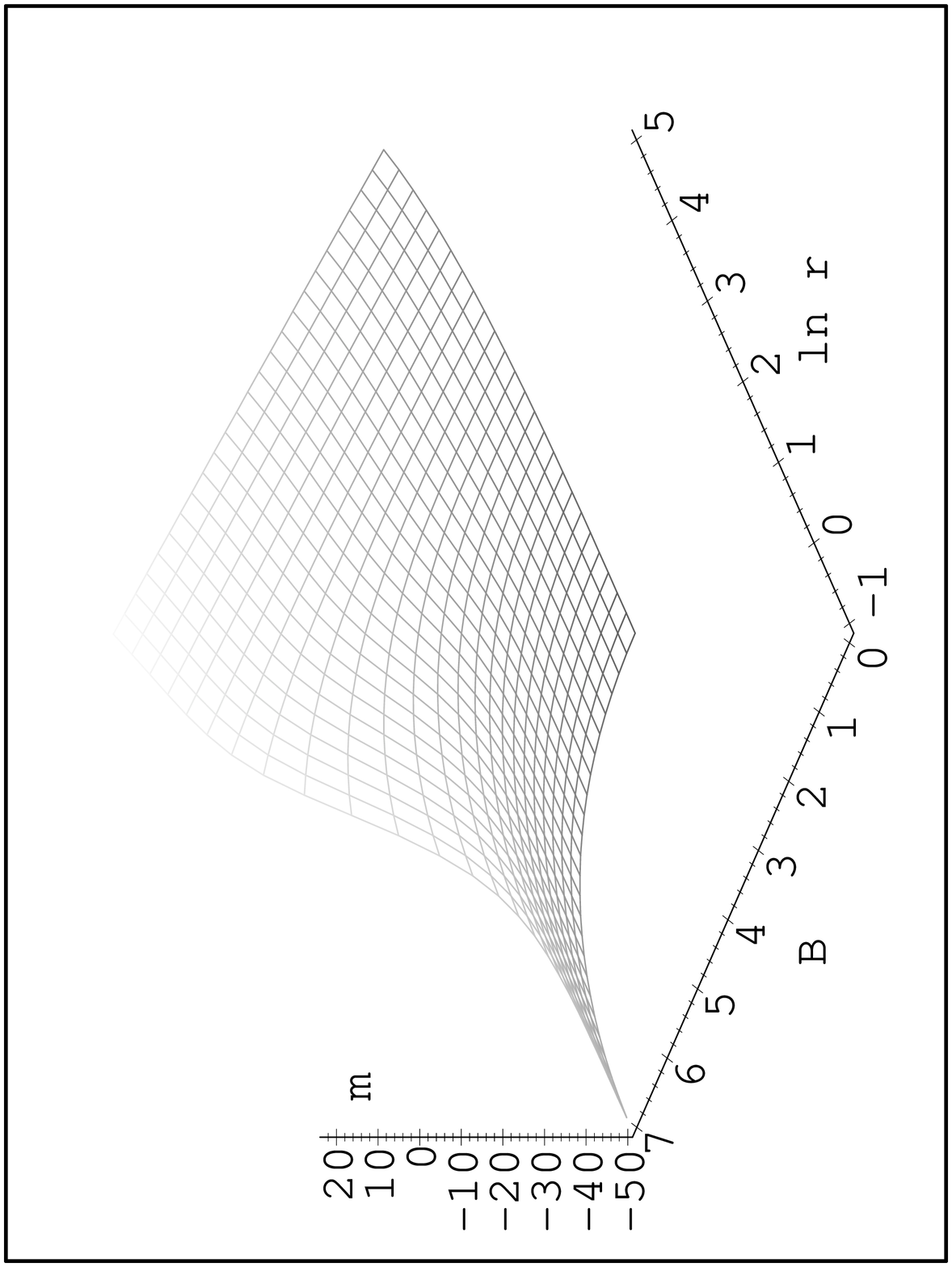}
\vspace*{2.3in}
\caption{(a) The mass function $m=m(\ln r)$ at parameter values $B=C=1$
takes negative values close to $r=0$. (b) The mass function in the
parameter region $0\leq B\leq 7$. The value of $r$ where
the mass becomes negative depends on the parameter $B$.
} \end{figure}
 
As we see from Fig. 3, the mass function vanishes at some radius
$\tilde r$ and takes negative values at $r<\tilde r$.
We show here that such a positive $\tilde r$
exists irrespective of the choice of the parameters $B$ and $C$.
By combining (\ref{rL}) and (\ref{mL}) with the condition $m=0$ we find
$\tilde r=2\Phi_B^2(\tilde L)/C\exp(\tilde L^2)>0$.
The constant $C$ has a simple scaling effect on the value of $\tilde r$.
Figure 3(b) shows that in the domain $(0,7)$ the negative mass
region can be extended by increasing $B$.
We emphasize that the energy conditions are still satisfied in
the negative mass regions.
The interpretation of the negative mass function is not immediate
and depends on the measuring procedure of the mass in
this asymptotically non-flat space-time.
 
At the end of this section we give the metric in double null and
radiation coordinates. This requires an additional integration.
We introduce a ``tortoise coordinate'' $r^*$ in the same manner it
can be introduced in the Schwarzschild space-time:
 
\begin{equation}
r^{*}=\int^{r}
\frac{dr^{\prime }}{\left[ f(r^{\prime })h(r^{\prime })\right]^{1/2}}
=\frac{\sqrt{2}}{C}\int^{L}\left[
-e^{y^{2}}+2y\Phi _{B}(y)\right] dy.  \label{tortoise}
\end{equation}
The radial null geodesics are $t=t_{0}+cr^{*}$, with $c=1$ for incoming and $
c=-1$ for outgoing geodesics.
 
Introducing the null coordinates $x^{\pm }=t\pm r^{*}$ the metric takes the
simple form:
\begin{equation}
ds^{2}=-h(x^{+},x^{-})dx^{+}dx^{-}+r^{2}(x^{+},x^{-})d\Omega ^{2}.
\label{metricDN}
\end{equation}
The radial null geodesics are now $x^{\pm }$=const. The functions $
h(x^{+},x^{-})=h(x^{+}-x^{-})$ and $r(x^{+},x^{-})=r(x^{+}-x^{-})$ are
contained in implicit form in Eqs. (\ref{g}),(\ref{tortoise}),
(\ref{rL}) and ( \ref{beL}).
 
Finally we cast our static, spherically symmetric solution of crossflowing
null dust in either the incoming or outgoing radiation coordinates $
(v=t+cr^{*},r,\theta ,\phi )$. These coordinates are like the
Eddington-Finkelstein coordinates for the Schwarzschild solution.
If the solution was asymptotically flat, they would be the
Bondi-Sachs coordinates.
\begin{equation}
ds^{2}=-\left(f\beta \right) ^{-1/2}dv\left[ \left(
\frac{f}{\beta} \right) ^{1/2}dv-2cdr\right] +r^{2}d\Omega ^{2}.
\label{metricEF} \end{equation}
Here $v=x^{\pm}$ for $c=\pm 1$. In these coordinates the
radial null geodesics are given by one of the equations $v$=const
and
 
\begin{equation}
v=const+2c\int^{r}\left( \frac{\beta \left( r^{\prime }\right) }{f\left(
r^{\prime }\right) }\right) ^{1/2}dr^{\prime }.  \label{geodEF}
\end{equation}
Now it is evident that there is no apparent horizon:
\begin{equation}
\frac{dr}{d\left( cv\right) }=\frac{1}{2}\left( \frac{f}{\beta }\right)
^{1/2}>0,
\end{equation}
thus no event horizon either. Thus the singularity in the origin
$r=0$ is naked. This is very similar to the naked singularity of
the {\sl negative mass} Schwarzschild solution \cite{Date}.
 
\section{Interpretation of the Solution}
 
\subsection{2D dilatonic model}
 
In this subsection we present a dilatonic model which in 4D has
the
interpretation of a spherically symmetric gravitational field in the
presence of two crossflowing null dust streams. This dilatonic
model emerges from the action
\begin{equation}
S=\int d^{2}x\sqrt{-g}\rho\left[ R\left[ g\right]
+\frac{1}{2}g^{\alpha \beta
}\nabla _{\alpha }\ln \rho \nabla _{\beta }\ln \rho +\frac{2}{\rho }\right] -
\frac{1}{2}\int d^{2}x\sqrt{-g}g^{\alpha \beta }\nabla _{\alpha }\varphi
\nabla _{\beta }\varphi\ .  \label{dilaton}
\end{equation}
The first term is the Einstein-Hilbert action reduced by spherical
symmetry (a surface term was dropped). The second term in
(\ref{dilaton}) represents a 2D massless scalar field in minimal
coupling. The minus sign assures that all 4D energy conditions are
satisfied. $\rho =r^{2}$ is the dilaton and $\varphi $ is the scalar field.
The conformal flatness of the $2\times 2$ metric $g_{\alpha \beta }=h\eta
_{\alpha \beta }$ is manifest in the corresponding 4D line element (\ref
{metricDN}). $\nabla $ and $R\left[ g\right] $ are the covariant derivative
and curvature scalar, respectively associated with the metric $g$. Although
the equations emerging from this model are quite similar to the
corresponding equations in the
Callan-Giddings-Harvey-Strominger (CGHS)
model\cite{CGHS}, they could not be exactly
integrated in general. For our purposes we take only the equation emerging
from variation of the scalar field, in double null coordinates:
\[
\varphi ,_{+-}=0
\]
Here commas denote derivatives. The equation has the D'Alembert solution
\begin{equation}
\varphi =\varphi ^{+}\left( x^{+}\right) +\varphi ^{-}\left( x^{-}\right) ,
\end{equation}
showing that the scalar field behaves like our matter source of crossflowing
null dust. The 4D interpretation of the particular solution with either
leftmoving or rightmoving matter is the Vaidya
solution\cite{Vaidya}. Recently Mikovi\'c has given a solution
\cite{Miko} for the case where both components are present, in the
form of a perturbative series in powers of the outgoing energy-momentum
component.
 
Our static solution (\ref {metricDN}) represents the first explicit
exact solution for this model, when none of the null dust
components are neglected.
 
\subsection{Anisotropic fluid}
 
Following Letelier \cite{Let1} we reinterpret the energy-momentum tensor
(\ref{en-mom}) as describing an anisotropic fluid with a pressure
component equaling its energy density:
\begin{eqnarray}
T_{ab} &=&\rho (U_{a}U_{b}+\chi _{a}\chi _{b})  \label{en-momAIF} \\
-U_{a}U^{a} &=&\chi _{a}\chi ^{a}=1\;,\qquad U_{a}\chi ^{a}=0\ .
\label{Uchi} \end{eqnarray}
 
A straightforward comparison with (\ref{en-mom}) gives
\begin{eqnarray}
\rho &=&\frac{\beta }{8\pi r^{2}},\qquad  \label{rho} \\
U^{a} &=&\frac{1}{\sqrt{2}}\left( v^{a}+u^{a}\right) =\frac{1}{\sqrt{h}}
\left( \frac{\partial }{\partial t}\right) ^{a}  \label{U} \\
\chi ^{a} &=&\frac{1}{\sqrt{2}}\left( v^{a}-u^{a}\right) =\sqrt{f}\left(
\frac{\partial }{\partial r}\right) ^{a}.  \label{chi}
\end{eqnarray}
 
Thus the solution represents an anisotropic fluid at rest with the energy
density $\rho $ and {\it radial} pressure $p=\rho .$ No tangential pressures
in the spheres $r$=const are present. The fluid is isotropic only
about a single point, the origin.
 
\subsection{Radiation atmosphere}
 
The solution can be interpreted as the outer region of a radiating star,
receiving radiation from the surrounding region either. If equilibrium is
achieved between the two components, we have the static solution (\ref
{metrictL}). This was the initial interpretation proposed by Date \cite{Date}
.
 
We write the energy-momentum tensor of the static solution in the double
null coordinate system $(x^{+},x^{-},\theta ,\phi )$. Inserting the null
covectors
\begin{equation}
v_{a}=\left( 0,-\sqrt{\frac{h}{2}},0,0\right) ,\qquad u_{a}=\left( -\sqrt{
\frac{h}{2}},0,0,0\right)  \label{vupm}
\end{equation}
in the covariant form of (\ref{en-mom}) and taking account of (\ref{g}) the
energy-momentum tensor becomes
 
\begin{equation}
T_{ab}dx^{a}dx^{b}=\frac{1}{16\pi r^{2}}\left(
dx^{+}dx^{+}+dx^{-}dx^{-}\right) ,  \label{en-momDN}
\end{equation}
a superposition of two cross-flowing null dust streams with equal
and {\it constant
}mass density functions. However one can freely rescale the null vectors $
v^{a}$ and $u^{a\text{ }}$ to have {\it arbitrary} mass density functions
either. After such a rescaling the mass density function of the incoming
null dust depends only on the outgoing coordinate $x^{+}$ and viceversa, a
property pertinent to the mass functions of the Vaidya solution
\cite{Vaidya}, characterized by the energy-momentum tensor
 
\begin{equation}
T_{ab}dx^{a}dx^{b}=\frac{c}{4\pi r^{2}}\frac{dM(V)}{dV}dVdV.
\label{en-momVa}
\end{equation}
$M(V)$ is the mass function and the coordinate $V$ is outgoing for incoming
radiation and incoming for outgoing radiation.
 
In the next section we will study the interior junction
with a static star, and the exterior junction with
incoming and outgoing Vaidya solutions.
 
\section{Junction Conditions}
 
\subsection{Matching with interior spherically symmetric
static solutions}
 
We discuss the junction with an interior solution with accent on the
anisotropic fluid interpretation given previously.
A similar treatment was given in \cite{Date}. Both our analysis
and the one in \cite{Date} reveals that the junction with a static
interior matter can be done without a regularizing thin shell. Our
treatment is more general, however. We formulate the junction
conditions for two generic static spherically symmetric space-times,
following the standard Darmois-Israel junction procedure
\cite{Darmois,Israel} and we establish a constraint on the matter
pressures implied by the matching conditions. An other improvement
over \cite{Date} is due to the fact that we dispose of the
exact solution (\ref{metrictL}), thus the explicit computation of
the the matching conditions with an arbitrary particular interior
becomes possible.
 
An orthonormal basis is given by the vectors $U$
and $\chi $ defined by Eqs. (\ref{U}) and (\ref{chi}) together
with the spacelike vectors
\begin{equation}
E_{{\bf 3}}=\frac{1}{r}\frac{\partial }{\partial \theta }\;,\qquad E_{{\bf 4}
}=\frac{1}{r\sin \theta }\frac{\partial }{\partial \phi }.  \label{E3E4}
\end{equation}
 
Any spherically symmetric static energy-momentum tensor has the form
\begin{equation}
T^{ab}=\rho U^{a}U^{b}+p_{1}\chi ^{a}\chi ^{b}+p_{2}\left( E_{{\bf 3}}^{a}E_{
{\bf 3}}^{b}+E_{{\bf 4}}^{a}E_{{\bf 4}}^{b}\right) .
\end{equation}
 
By inserting $p_{i}=\alpha _{i}/8\pi r^{2}$ and $\rho $ in the form (\ref
{rho}) in the Einstein equations for the metric (\ref{metric}) we find
\begin{eqnarray}
r\frac{df}{dr} &=&-\beta -f+1  \nonumber \\
rf\frac{dh}{dr} &=&h(\alpha _{1}-f+1)  \label{systfluid} \\
2rf\frac{d\alpha _{1}}{dr} &=&-\alpha _{1}^{2}+\alpha _{1}(-\beta
+f-1)+\beta (f-1)+4\alpha _{2}f.  \nonumber
\end{eqnarray}
 
The induced metric of the surface $r$=const is
\begin{equation}
ds_\Sigma^2=-hdt^2+r^2d\Omega^2\ .
\end{equation}
Without loss of generality we can choose the time coordinates
in both static space-times such that they are continuous on the
junction. Then the continuity of the first fundamental form requires
the metric function $h$ to be continuous.
 
The extrinsic curvature of the junction surface $r=const$ is
defined as
\begin{equation}
K_{ab}=\left( \delta _{a}^{c}-\chi ^{c}\chi _{a}\right) \left( \delta
_{b}^{d}-\chi ^{d}\chi _{b}\right) \nabla _{c}\chi _{d}\ .
\end{equation}
The nonvanishing components are:
\begin{equation}
K_{tt}=\frac{h}{2r\sqrt{f}}(-\alpha _{1}+f-1)\;
,\qquad K_{\theta }^{\theta }=K_{\phi
}^{\phi}=\frac{\sqrt{f}}{r}\ . \end{equation}
In the above expressions the derivatives were eliminated by use of
Eq. (\ref {systfluid}).
 
Continuity of the extrinsic curvature across the junction hypersurface
$r=r_{1}$ is achieved, provided that the metric and the function
$\alpha _{1}$ (thus also $p_{1}$) are continuous. We have proved
the following result:
 
{\it Any two spherically symmetric static solutions can be matched
along hypersurfaces} $r=const$ {\it provided the radial pressures
are continuous.} This is similar to the theorem given by Fayos,
Ja\'en, Llanta and Senovilla \cite{Seno} for the matching of the
Vaidya solution with a generic spherically symmetric solution
along timelike hypersurfaces. A generic discussion on matching
spherically symmetric space-times along thin spherical timelike
shells can be found in\cite{Lake}.
 
For the double null dust solution $\alpha _{1}=\beta$.
In consequence the interior fluid should have a radial pressure equal
to the energy density (\ref{rho}) of the double null dust solution
on the junction. However, no conditions on the pressures tangent to the
spheres emerge.
 
We see from Eqs. (\ref{beL}), (\ref{mL}) and (\ref{rho}) that the
integration constants $B$ and $C$ appear in the radial
pressure and mass function of the static double null dust
solution. Continuity of these functions on the junction fixes
the value of the constants, once the interior solution is
chosen.
For a realistic star, the mass $M_{1}$ should be positive. This
implies a lower boundary for the possible values of $r_{1}$, as
follows from Fig. 3(a).
 
Let us illustrate the junction with the interior Schwarzschild
solution\cite{Schw}
$ds^2=-\left(a-b\sqrt{1-r^2/R^2}\right)^2dt^2
     +\left(1-r^2/R^2\right)^{-1}dr^2+r^2d\Omega^2 $,
with the energy density $\rho_S=3/R^2$=const and pressure $p_S$
given by
\begin{equation}
8\pi p_S=\frac{3b\sqrt{1-\frac{r^2}{R^2}}-a}
              {R^2\left(a-b\sqrt{1-\frac{r^2}{R^2}}\right)}\ .
\end{equation}
Several relations among the Schwarzschild parameters $a, b, R$,
radius of the star $r_1$ and the parameters $B$ and $C$
emerge from the junction conditions:
\begin{equation}
2m_1=\frac{r_1^3}{R^2}
\ ,\qquad
2a=\frac{3}{\sqrt{\beta_1}}+\frac{R^2}{r_1^2}\sqrt{\beta_1}
\ ,\qquad
2b\sqrt{1-\frac{r_1^2}{R^2}}
=\frac{1}{\sqrt{\beta_1}}+\frac{R^2}{r_1^2}\sqrt{\beta_1}
\ .
\label{params}
\end{equation}
We have denoted by $\beta_1$ and $m_1$ the values of the functions
$\beta$ and $m$ at the junction $r=r_1$. The first two relations
(\ref{params}) determine the constant $B$ and the value of the
radial coordinate $L_1$ at the junction in terms of $r_1/R$ and $a$,
when Eqs. (\ref{rL}), (\ref{beL}) and (\ref{mL}) are inserted.
Eliminating
$\beta_1$ from the last two relations of (\ref{params}), a constraint
$b=b(a,r_1/R)$ on the possible values of characteristics of the interior
emerges. Finally (\ref{rL}) implies $C=C(a,r_1/R,r_1)$.
 
In the light of the above relations we see that after choosing
some value for $C$ (a scale), the constant $B$ is determined
exclusively by the radius and density (or mass) of the star.
 
\subsection{Matching with Vaidya solutions}
 
In this subsection we study the junctions with the incoming and
outgoing Vaidya
solutions, which are at the exterior of the static double null dust
solution. There are only three points (in fact spheres) in common with
exterior Schwarzschild regions (Fig. 4), and the matching can be
done without introducing regularizing thin shells.
 
The high-frequency approximation to the unidirectional radial flow of
spherically symmetric unpolarized radiation, characterized by the
energy-momentum
tensor (\ref {en-momVa}) is represented by the Vaidya
solution\cite{Vaidya}:
\begin{equation}
ds^{2}=-dV\left[ \left( 1-\frac{2M(V)}{r}\right) dV-2cdr\right]
+r^{2}d\Omega ^{2}.  \label{Va1}
\end{equation}
The radial null geodesics are the lines $V$=const and the curves
given by \begin{equation}
\frac{dr}{dV}=\frac{c}{2}\left( 1-\frac{2M(V)}{r}\right) .  \label{geod}
\end{equation}
 
We would like to match our static solution with incoming and outgoing
Vaidya solutions along the outgoing, respectively incoming radial null
geodesics, e.g. along lines of constant incoming, respectively outgoing
coordinate (Fig 4.). These are given by Eq. (\ref{geod}) in the
Vaidya solution and by Eq. (\ref{geodEF}) in the static solution.
 
A convenient formalism for matching solutions along null surfaces, which
does not require coordinates that match continuously on the shell, was
developed by Barrab\`{e}s and Israel\cite{BI}. Their discussion on the
particular case of spherical symmetry requires the metric in both
space-times written in the form:
\begin{equation}
ds^{2}=-e^{\psi }dz\left[ fe^{\psi }dz-2dr\right] +r^{2}d\Omega ^{2}.
\label{gensphe}
\end{equation}
Here $f$ and $\psi$ depend on both $r$ and $z$. The null coordinate
$z$ is always outgoing but it can be either increasing or decreasing
with time. Then the junction is done along the null hypersurfaces
$z$=const.
 
To apply this formalism it would be desirable to express the Vaidya solution
in the other set of radiation coordinates $[U(V,r),r,\theta ,\phi
]$ in
which Eq. (\ref{geod}) takes the form $U$=const. However, this is
possible only
when double null coordinates are found. It was demonstrated in \cite{WLake}
that double null coordinates exist when the mass is a {\it linear }or {\it
exponential} function of the advanced or retarded time $V$. We
argue in what follows that the mass function of that particular
Vaidya solution, which
is matching continuously to the static solution has a more complicated
dependence. A further inconvenience is that there is no obvious choice for
the intrinsic coordinates in terms of the space-time coordinates in the null
junction hypersurface. For these reasons we proceed as follows. First
we find
coordinates that match continuously on the junction and in terms of which
the metric is continuous. Then we employ the Barrab\`{e}s-Israel
junction formalism to
find the distributional stress-energy tensor on the junction.
 
\begin{figure}[tb]
\epsfysize=8cm
\centerline{\hfill\epsfbox{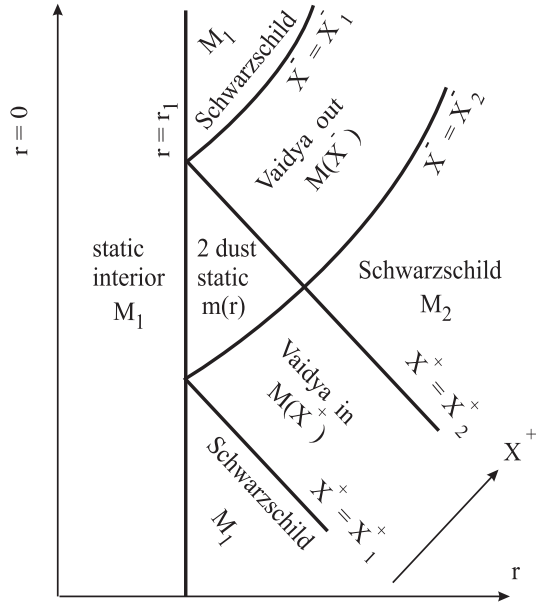}\hfill}
\vspace*{.2in}
\caption{The intersection of two crossflowing null dust streams
with mass functions
$M(V)$ is the static solution characterized by the mass function $m(r)$.
Here $V=X^{\pm }$ stands for $c=\pm 1$ and $X^{\pm }$ are the advanced
(retarded) time in the Vaidya solution. The 2-dust solution touches in three
points exterior Schwarzschild regions with masses $M_1$ and $M_2$. At the
interior junction there is a static fluid representing a star with mass $M_1$
. The mass functions $M(V)$ of the Vaidya regions change monotonously from $
M_1$ to $M_2$ . }
\end{figure}
 
As a radial variable of both metrics we choose $L$ by extending the
expression (\ref{rL}) of $r=r(L)$ to the Vaidya regions too. An appropriate
null coordinate $z$ is defined by the values of $L=L_{j}$ on the
junction. In these coordinates the junction hypersurfaces are
characterized by the null geodesic equations $z=L$ in both space-times.
We proceed in deriving the expressions of the coordinates $v$ and $
V$ and of the mass function $M$ in terms of the coordinate $z$.
 
Identifying the corresponding part of Eq. (\ref{en-momDN}) with
Eq. (\ref{en-momVa}
), we find the relation between the null coordinates $V$ and $v$, valid on
the junction :
\begin{equation}
dv=2\left( c\frac{dM(V)}{dV}\right) ^{1/2}dV.  \label{vV}
\end{equation}
Then we extend this relation over both the static and the Vaidya regions.
 
The null geodesic equation (\ref{geodEF}) of the static solution, evaluated
on the junction (where $z=L$) gives
\begin{equation}
\frac{dz}{dV}=\frac{c}{2r^{2}}\frac{dM}{dz}  \label{zV}
\end{equation}
Inserting Eq. (\ref{zV}) in the null geodesic equation
(\ref{geod})
of the Vaidya solution we find its mass function in terms of
the new null coordinate $z$:
\begin{equation}
2CM(z)=-e^{z^{2}}+2z\Phi _{B}(z)-2e^{-z^{2}}\Phi _{B}^{2}(z).
\label{Mz} \end{equation}
Thus the mass function is continuous at the junction. The geodesic equation
(\ref{zV}) together with (\ref{Mz}) gives the relation between the null
coordinates $V$ and $z$:
\begin{equation}
V\left( z\right) =const+\frac{2c}{C}\int^{z}\frac{e^{y^{2}}\left[
-e^{y^{2}}+2y\Phi _{B}(y)\right] dy}{\Phi _{B}(y)}.  \label{Vz}
\end{equation}
The relations (\ref{Mz}) and (\ref{Vz}) contain in implicit form
the dependence of the mass function $M$ of the Vaidya solution on
the radiation coordinate $V$. Despite the complicated dependence
$M(V)$ it is straightforward to check that $M$ satisfies the
required monotonicity condition:
\begin{equation}
2\frac{dM}{d(cV)}=\frac{\Phi_B^2(z)}{e^{2z^{2}}}>0\ .
\end{equation}
Finally Eqs. (\ref{vV}) and (\ref{Vz}) give $v=v\left( z\right)$:
\begin{equation}
v\left( z\right) =const+\frac{2\sqrt{2}c}{C}\int^{z}\left[
-e^{y^{2}}+2y\Phi _{B}(y)\right] dy.  \label{vz}
\end{equation}
 
Now we express both metrics in terms of the coordinates
$(z,r(L),\theta,\phi )$. They take the form (\ref{gensphe}) with
\begin{eqnarray}
f &=&1-\frac{2m\left( L\right) }{r\left( L\right) }\;,\qquad
e^{\psi }=\frac{2}{C}\frac{e^{L^{2}}}{\Phi _{B}(L)}
          \left[ -e^{z^{2}}+2z\Phi _{B}(z)\right]
\label{fpsi} \\
f_{V} &=&1-\frac{2M\left( z\right) }{r\left( L\right) }\;,\qquad
e^{\psi_{V}}=\frac{2}{C}\frac{e^{z^{2}}}{\Phi _{B}(z)}
             \left[ -e^{z^{2}}+2z\Phi_{B}(z)\right] .
\label{fpsiVa}
\end{eqnarray}
Here the index $V$ refers to the Vaidya solution and the expressions
$r\left( L\right) ,m\left( L\right) $ and $M\left( z\right) $ are
given by Eqs. (\ref{rL}), (\ref{mL}) and (\ref{Mz}), respectively.
We
have completed the task of writing both metrics in coordinates which
are continuous on the junction and in terms of which the metric is
continuous.
 
It is immediate to check the continuity of the induced metric
given by $ds^2_\Sigma=r^2d\Omega^2$. The other junction
condition is a somewhat subtle issue as the conventional
extrinsic curvature tensor for null hypersurfaces carries no
transversal information.
 
We define a pseudo-orthonormal basis \cite{HE} $(n,N,E_{{\bf 3}},E_{{\bf 4}
}) $, where $E_{{\bf 3}}$ and $E_{{\bf 4}}$ are given in
Eq. (\ref{E3E4}) and
\begin{equation}
n=e^{-\psi }\frac{\partial }{\partial z}
 +\frac{f}{2}\frac{\partial }{\partial r}\;,\qquad
N=-\frac{\partial }{\partial r}\;.
\end{equation}
The vector $n^a$ is orthogonal (and also tangent) to the
hypersurfaces
$fe^{\psi}dz-2dr=0$, along which the two space-times are glued together.
The vector $N^a$ is the other radial null vector, transversal to
these surfaces. They are related to the previously introduced vectors
$v^a$ and $u^a$ as follows. For the
junction with the incoming Vaidya region $v^a=\left( 2/f\right)
^{1/2}n^a$ and $
u^a=\left( f/2\right) ^{1/2}N^a,$ while for the junction with the
outgoing
Vaidya region $u^a=-\left( 2/f\right) ^{1/2}n^a$ and $v^a=-\left(
f/2\right) ^{1/2}N^a$.
 
The projector to these null hypersurfaces with tangent space spanned by
$E_{{\bf 3}},E_{{\bf 4}}$ and $n$ is
\begin{equation}
P_{b}^{a}=E_{{\bf 3}}^{a}E_{b}^{{\bf 3}}+E_{{\bf 4}}^{a}E_{b}^{{\bf 4}
}-n^{a}N_{b}=\delta _{b}^{a}+N^{a}n_{b}\ .
\end{equation}
Following \cite{BI} we define the transverse or oblique extrinsic
curvature tensor
with the aid of the transverse vector $N$:
\begin{equation} \label{Kab}
K_{ab}=P_{a}^{c}P_{b}^{d}\nabla _{c}N_{d} \ .
\end{equation}
The only nonvanishing components of the transverse extrinsic curvature
tensor\footnote{The components of the extrinsic curvature tensor defined
in \cite{BI} are found from $K_{ab}$
by contracting with the three basis vectors
$E_{{\bf 3}},E_{{\bf 4}}$ and $n$ tangent to the hypersurface.}
 are
\begin{equation}
K_{zz}=\frac{1}{2}e^{\psi }\partial _{r}\left( fe^{\psi }\right)
=\frac{C}{2
}e^{\psi }\frac{dL}{d\left( Cr\right) }\partial _{L}\left( fe^{\psi }\right)
\;,\qquad K_{\theta }^{\theta }=K_{\phi }^{\phi }=-\frac{1}{r}.
\end{equation}
Thus {\it when matching any two metrics of the form}
(\ref{gensphe})
{\it on the null hypersurfaces} $fe^{\psi }dz-2dr=0$,
{\it the jump in the extrinsic curvature is given by the jump in}
$\partial _{r}\left( fe^{\psi }\right) $,
{\it provided the metric is continuous across the junction.}
 
A straightforward computation employing (\ref{fpsi}) and (\ref{fpsiVa})
gives on the junction hypersurfaces of the static double null dust
region with the Vaidya regions
\begin{equation}
\partial _{L}\left( fe^{\psi }\right)
=\partial _{L}\left( f_{V}e^{\psi_{V}}\right) =
\frac
{4e^{L_{j}^{2}}\left[-e^{L_{j}^{2}}+2L_{j}\Phi_{B}(L_{j})
                     -e^{-L_{j}^{2}}\Phi_{B}^{2}(L_{j})\right]}
{C\left[-e^{L_{j}^{2}}+2L_{j}\Phi_{B}(L_{j})\right]}
\ .
\end{equation}
 
In conclusion the extrinsic curvature is also continuous on the junction.
There is no need for a thin regularizing shell separating the two domains of
the space-time, in contrast to the exterior junction proposed in \cite{Date}.
 
\section{Concluding Remarks}
We have integrated the Einstein equations in the presence of
crossflowing null dust under the assumptions of spherical
symmetry and
staticity and analyzed various aspects related to the properties
of the emerging exact solution (\ref{metrictL}). The solution
has dilatonic gravity connections and it can be reinterpreted
as an anisotropic fluid with radial pressure equal to its energy
density and no pressures along the spheres $r$=const. This can be
a radiation atmosphere for a star with its radial pressure equal
to the energy density of the athmosphere on its surface. No
constraint on the pressures along the spherical junction surface
was found. On the exterior, the study of the matching conditions
with the Vaidya solution revealed no thin shells on the junction.
 
As a byproduct, we have derived general conditions for the
junction of two spherically symmetric solutions. Matching of two
static space-times (\ref{metric}) along $r$=const hypersurfaces is
assured by the continuity of the metric functions and of the radial
pressure. Matching of generic spherically symmetric space-times
(\ref{gensphe}) along the null hypersurfaces $fe^{\psi}dz-2dr=0$
is possible whenever the metric and
$\partial _{r}\left( fe^{\psi }\right) $ are continuous.
 
The negativity of the mass function in some neighbourhood of the
$r=0$ singularity raises the possibility of matching this solution
to a negative mass core. This may be difficult due to the nontrivial
topology of the known negative mass black hole solutions\cite{Mann}.
Despite the lack of experimental evidence for negative mass objects,
presumably of quantum origin\cite{Date}, their microlensing
effect\cite{Cramer} on radiation
from Active Galactic Nuclei was shown to produce features similar to
some observed Gamma Ray Bursts\cite{Torres}.
 
Equally interesting would be to introduce dynamics
in the picture by an interior matching with a collapsing star.
We defer this topic to a forthcoming study.
 
An intriguing open question remains
whether exact solutions describing the collision of spherically
symmetric null dust streams, which have not reached equilibrium,
can be found.
 
{\sl Note added.} After the submission of this paper a relevant
work in the subject
was published by Kramer \cite{Kramer3}. The solution
presented there is the particular case of the metric
(\ref{metrictL}) with the parameter values $B=0$ and $C=\sqrt{e}$.
Our metric functions $h$ and $f$, radial variable $L$ and
transcendental function $-e^{L^2}+2L\Phi_B$ correspond to
$e^{2\nu},\ e^{-2\lambda},\ \sqrt{1+2\xi}$ and $
\sqrt{e(1+2\xi)}J$, respectively of this paper, when the
parameters values $B=0$ and $C=\sqrt{e}$ are chosen. Keeping the
parameters arbitrary enabled us in Sec. IV A to match the
solution (\ref{metrictL})
with an interior Schwarzschild solution with {\sl arbitrary} mass and
radius. In contrast with our analysis relying on the
Barrab\`{e}s-Israel
matching procedure, in \cite{Kramer3} the junction with
the Vaidya space-time was discussed by imposing the continuity of the
four-metric on the junction.
 
\section{Acknowledgments}
 
The author is grateful to Ji\v {r}\'{\i} Bi\v {c}\'{a}k, Gyula Fodor, Karel
Kucha\v {r} and Zolt\'{a}n Perj\'{e}s for discussions on the subject and
helpful references.
This work has been supported by OTKA no. W015087 and D23744 grants. The
algebraic packages REDUCE and MAPLEV were used for checking
computations and numerical plots.

\end{document}